\documentclass[namedreferences]{SolarPhysics}
\usepackage[optionalrh,natbib]{spr-sola-addons}
\usepackage{epsfig}
\usepackage{graphicx}
\usepackage{courier}
\usepackage{color}
\usepackage{url}
\usepackage{setspace}

\newcommand{\adv}{    {\it Adv. Space Res.}}

\newcommand{\aapr}{   {\it Astron. Astrophys. Rev.}}

\newcommand{\solphys}{{\it Solar Phys.}}

\newcommand{\ssr}{    {\it Space Sci. Rev.}}

\begin{document}
\begin{article}
\begin{opening}
\title{%
Origin and Use of the Laplace Distribution in Daily Sunspot Numbers
}
\author{%
P.L.~\surname{Noble}$^{1}$ \sep
M.S.~\surname{Wheatland}$^{1}$ \sep
}
\runningauthor{Noble, P.~L., Wheatland, M.~S.}
\runningtitle{Origin and Use of the Laplace Distribution in Daily Sunspot Numbers}
\institute{%
$^{1}$Sydney Institute for Astronomy, \\
School of Physics, \\
The University of Sydney, \\
Sydney NSW 2006, \\
Australia \\
email: \url{p.noble@physics.usyd.edu.au}
}

\begin{abstract}
Recently Pop ({\it Solar Phys.} {\bf 276}, 351, 2012) identified a Laplace (or double 
exponential) distribution in the number of days with a given absolute value in the change over a 
day, in sunspot number, for days on which the sunspot number does change. We show this 
phenomenological rule has a physical origin attributable to sunspot formation, evolution, and 
decay, rather than being due to the changes in sunspot number caused by groups rotating onto 
and off the visible disc. We also demonstrate a simple method to simulate daily sunspot 
numbers over a solar cycle using the \cite{2012SoPh..276..351P} result, together with a model 
for the cycle variation in the mean sunspot number. The procedure is applied to three recent 
solar cycles. We check that the simulated sunspot numbers reproduce the observed distribution 
of daily changes over those cycles.
\end{abstract}

\keywords{Solar Cycle, Models; Sunspots, Statistics}
\end{opening}

\section{Introduction}
\label{Section:Introduction}

Sunspots have been studied scientifically since the invention of the telescope, and reliable daily 
sunspot number records are available from the early 1800s\footnote{Sunspot data are compiled 
by the US National Geophysical Data Center (NGDC), and are available at 
\url{http://www.ngdc.noaa.gov/stp/solar/ssndata.html}. The NGDC data are used throughout in 
this paper.}. The accepted quantity to characterise solar activity is the international sunspot 
number
\begin{equation}\label{eq:issn}
s = k (10g + n),
\end{equation}
where $n$ is the number of individual sunspots, $g$ is the number of sunspot groups, and $k$ is 
a correction factor \citep{1977ASSL...69.....B}. The sunspot number changes in a secular or 
long-term fashion with the semi-regular 11-year sunspot cycle, driven by an internal magnetic 
dynamo which generates the magnetic field \citep{2002RSPTA.360.2741T}. The secular 
change in sunspot number exhibits apparent randomness, as evidenced by the extensive 
literature describing the smoothed daily sunspot number (typically a monthly average) as a 
stochastic, or chaotic time series. In particular, the peak and timing of the different solar cycles 
shows considerable variation \citep{2010LRSP....7....6P}.

Sunspot numbers also vary on shorter time scales, in particular daily, as a result of the 
complicated local processes associated with sunspot formation, evolution, and decay. The 
short-timescale variation produces large excursions in sunspot number, up to 100 per day in 
extreme cases \citep{2011ApJ...732....5N}. The large day to day variations are caused by the 
rapid evolution of magnetic structures, and the sudden appearance/disappearance of large active 
regions. These rapid developments can have important consequences for the space weather 
experienced on Earth \citep{2008sswe.rept.....C}.

Recently, it was demonstrated that the change $\Delta s$ in the daily sunspot number (for days 
on which the number does change) follows a Laplace, or double exponential distribution 
\citep{2012SoPh..276..351P}. Including the case of days with no changes, 
\cite{2012SoPh..276..351P} modelled the distribution $f(\Delta s)$ of the change in sunspot 
number with the functional form
\begin{equation}\label{eq:pop_law}
f(\Delta s) = A\mathbb{I}(\Delta s<0)\exp(\Delta s/B) +
 A\mathbb{I}(\Delta s>0)\exp(-\Delta s/B)+C\delta(\Delta s),
\end{equation}
where $A, \, B$, and $C$ are constants, and where $\mathbb{I}(x)$ is the indicator function 
defined by $\mathbb{I}(x)=1$ for $x$ true, and $\mathbb{I}(x)=0$ for $x$ false. The 
parameter $C$ determines the fraction of zero changes, and $B$ is the mean absolute change for 
days on which the number does change. Normalisation of Equation~(\ref{eq:pop_law}) 
requires $A = (1-C)/2B$.

\begin{figure}[ht!]
\centerline{\includegraphics[height=4.0in]{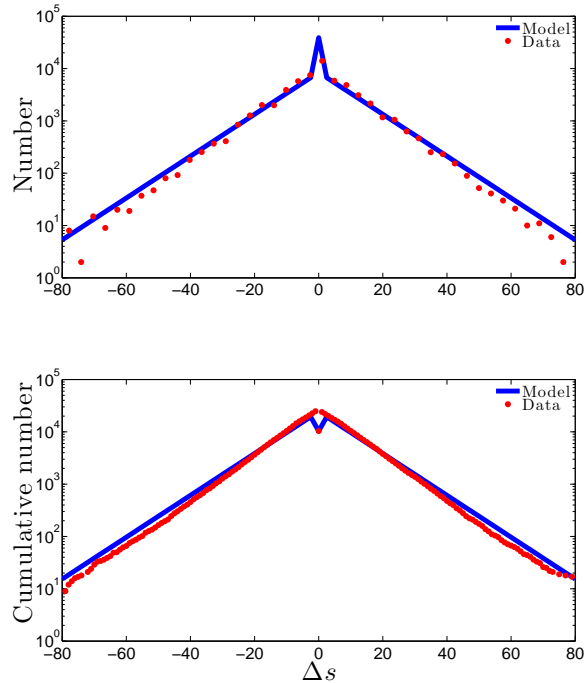}
}
\caption{%
Histograms showing the observed numbers of daily changes in sunspot number for 1850--2011 
(red points), and the model distribution for the changes defined by 
Equation~(\ref{eq:pop_law}). The upper panel shows the number of days with a given change, 
for positive and negative changes, and the lower panel shows the corresponding cumulative 
number.
}
\label{fig:f1}
\end{figure}

Figure~\ref{fig:f1} shows the observed distribution for the daily sunspot number using the 
NGDC data for 1850--2011, and illustrates the exponential form identified by 
\cite{2012SoPh..276..351P}. The top panel is a histogram of daily changes $\Delta s$. The 
bottom panel shows the cumulative number of changes greater than $\Delta s_i$, for positive 
changes:
\begin{equation}
N(\Delta s \geq \Delta s_i)=\sum_i \mathbb{I}(\Delta s \geq \Delta s_i >0),
\end{equation}
and the cumulative number less than $\Delta s_i$ for negative changes:
\begin{equation}
N(\Delta s \leq \Delta s_i) = \sum_i \mathbb{I}(\Delta s \leq \Delta s_i<0).
\end{equation}
Both panels use a logarithmic scaling on the vertical axis. The red dots in the panels show the 
data, and the blue curves show the model distribution defined by Equation (\ref{eq:pop_law}), 
with parameters $B$ and $C$ estimated from the data using maximum likelihood 
\citep{Eliason1993}. The adherence to the exponential form in Figure \ref{fig:f1} is striking. 
There are a large number of days with no change in sunspot number, and 
\cite{2012SoPh..276..351P} refers to $\Delta s=0$ as a ``special state''. The distribution is 
remarkably symmetric about $\Delta s=0$. \cite{2012SoPh..276..351P} also investigated the 
solar cycle dependence of the observed distribution, and found that the exponential rule in 
absolute changes is most closely adhered to over whole cycles, with the greatest departure near 
minima of cycles.

It is surprising that the Laplace distribution in the change in daily sunspot number was not 
identified and discussed in the literature earlier. The behaviour was previously noted in 
smoothed data \citep{2009IJMPB..23.5609L}, and an approximate exponential dependence in 
the distribution of overall sunspot numbers, related to Equation~(\ref{eq:pop_law}), was also 
commented on \cite{2011ApJ...732....5N}. However, \cite{2012SoPh..276..351P} showed that 
the adherence of the observed changes in daily sunspot number to the Laplace distribution 
defined by Equation (\ref{eq:pop_law}) is much stricter than the approximate exponential 
distribution of overall sunspot number. The Laplace distribution (Equation~(\ref{eq:pop_law})) 
represents a newly identified phenomenological rule describing the way in which the daily 
sunspot number varies, which should have application for modelling and prediction. The origin 
of the distribution is not obvious. Changes in sunspot number occur daily due to sunspot group 
formation, evolution and decay, the appearance/disappearance of spots and spot splitting 
(referred to here as spot evolution), and also due to sunspot groups and individual spots rotating 
onto and off the visible disc. If the law is due to spot and group evolution, then the 
phenomenological rule must have a physical origin.

In this paper we demonstrate that the Laplace distribution of changes in sunspot number is 
caused by sunspots forming, evolving, and decaying, and is not a result of rotation on and off 
the disc. We then also present a simple Monte Carlo method, based on 
Equation~(\ref{eq:pop_law}) and some additional assumptions, for simulating sunspot 
numbers over solar cycles.

\section{Origin of the Laplace Distribution}
\label{Section1}
To investigate the origin of the observed distribution in changes in sunspot number we use 
reports of sunspot groups on the Sun, for 1981-2011, compiled by US National Oceanic and 
Atmospheric Administration (USAF/NOAA).\footnote{Data are available at 
\url{http://ngdc.noaa.gov/stp/solar/sunspotregionsdata.html}.} The daily change in sunspot 
number
\begin{equation}
\label{eq:changeins}
\Delta s = k \left(10 \Delta g + \Delta n \right)
\end{equation}
can be decomposed into changes due to rotation of regions and spots onto and off the disc 
$\Delta s_{\rm r}$, and changes due to group and sunspot evolution $\Delta s_{\rm e}$:
\begin{equation}
\Delta s = \Delta s_{\rm r} + \Delta s_{\rm e}.
\end{equation}
Similarly the terms $\Delta g$ and $\Delta n$ on the right hand side of Equation 
(\ref{eq:changeins}) can be decomposed in this way. To approximate the change due to 
rotation $\Delta s_{\rm r}$ we assume that active regions first appearing on the disc within 
$180/14\approx 13^{\circ}$ of the eastern limb arrive due to rotation within a day, and regions 
last observed on the disc within $13^{\circ}$ of the western limb disappear due to rotation 
within a day. The factor of 14 days is the approximate time to rotate across the disc 
\citep{1990ApJ...351..309S}.

\begin{figure}[ht!]
\centerline{\includegraphics[height=2.5in,width=4in]{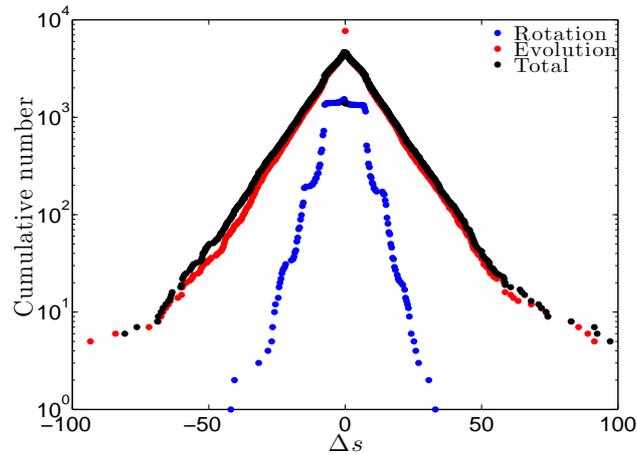}
}
\caption{%
The cumulative distribution of the changes in daily sunspot number due to rotation of sunspot 
regions onto and off the disc (blue), the change due to evolution of regions (red), and the total 
change (black).
}
\label{fig:f2}
\end{figure}

Figure~\ref{fig:f2} shows the result of the analysis of the data. The figure presents the 
cumulative distribution of the total change in sunspot number $\Delta s$ using the complete 
data set (black), the change due to rotation of spots and groups onto and off the disc $\Delta 
s_{\rm r}$ (blue), and the changes due to the evolution of spots and groups $\Delta s_{\rm 
e}$ (red). The distributions of the total change $\Delta s$ and change due to evolution $\Delta 
s_{\rm e}$ are very similar, and both clearly show the Laplace distribution of 
Equation~(\ref{eq:pop_law}). The distribution of $\Delta s_{\rm r}$ exhibits significant falls 
in number at $\Delta s_{\rm r} = \pm 10k, \pm 20k,...$, which may be attributed to entire 
sunspot groups rotating onto and off the disc (groups are weighted with a factor of 10 in the 
definition of the international sunspot number -- see Equation (\ref{eq:issn})).

An important feature of Figure~\ref{fig:f2} is that the number of changes due to rotation onto 
and off the visible disc is at least an order of magnitude smaller than the number of changes due 
to evolution. There are too few changes associated with active regions rotating onto and off the 
visible disc for this to influence the overall distribution of changes in sunspot number. It is very 
unlikely that the observed double exponential distribution of changes, for days on which the 
sunspot number does change, is influenced by the small number of active regions which rotate 
onto or off the disc in a day.

\begin{figure}[ht!]
\centerline{\includegraphics[height=2.5in,width=4in]{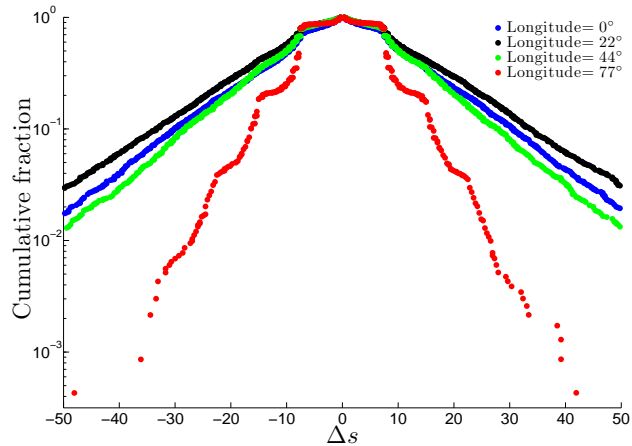}
}
\caption{%
The cumulative distribution of daily change sunspot number calculated in four 
26$^\circ$ degree strips. The daily changes in sunspot number in the strip from $-13^\circ$ to 
$13^\circ$ are shown in blue, the changes in the strip from $-35^\circ$ to $-22^\circ$ and 
from $22^\circ$ to $35^\circ$ are shown in black, the change in the strip from $-57^\circ$ to 
$-44^\circ$ and from $44^\circ$ to $57^\circ$ are in green, and changes in the strip from 
$-90^\circ$ to $-77^\circ$ and from $77^\circ$ to $90^\circ$ are shown in red.
}
\label{fig:f2x}
\end{figure}

It is possible that the observed Laplace distribution arises from an averaging of locally 
non-exponential distributions across the disk. To test this we also calculate the cumulative 
fraction of changes occurring in particular 26$^\circ$ strips on the disc. In other words, we 
identify the active regions in the strips on a given day, and calculate the sunspot number due to 
these regions only. On the next day we identify the regions in the strips, and again count the 
sunspot number due to these regions. The difference between these two counts is the change in 
sunspot number in a day occurring in the particular strips. This count includes both changes due 
to rotations into and out of the strips, and changes due to evolution in the strips. 
Figure~\ref{fig:f2x} shows the results. Changes occurring in the region 
$(-13^\circ,0^\circ)$ and $(0,13^\circ)$ are in blue, changes occurring in the region 
$(-35^\circ,-22^\circ)$ and $(22,35^\circ)$ are in black, changes occurring in the region 
$(-57^\circ,-44^\circ)$ and $(44,57^\circ)$ are in green, and changes occurring at the limbs 
({\it i.e.} $(-90^\circ,-77^\circ)$ and $(77,90^\circ)$) are in red. Figure~\ref{fig:f2x} shows 
that changes associated with regions in a restricted range of longitudes reproduce the observed 
Laplace distribution. The exception is the distribution of changes at each limb, which shows 
large falls in probability due to changes in the number of groups in the strips, and a significantly 
smaller proportion of large changes in sunspot number. Presumably these differences arise due 
to the difficulty of accurately observing sunspot regions at the limb, and in particular, resolving 
the number of individual spots in a group. This may explain why this particular distribution is 
dominated by changes in the number of groups.

Figure~\ref{fig:f2x} shows that the observed Laplace distribution in total changes does not 
arise from averaging over non-exponential local distributions in particular strips. The 
conclusion is that the Laplace distribution form is representative of changes in daily sunspot 
number due to the local evolution of sunspot groups on the Sun. It reflects the physical 
processes occurring.

\section{Daily Change in Sunspot Number}
\label{Section:Section2}
\subsection{A Conditional Distribution for Daily Change in Sunspot Number}
\label{Section:Section3.1}
\cite{2012SoPh..276..351P} investigated the daily change in sunspot number $\Delta s$ using 
data over complete cycles, which involve days with a range of different initial values $s$ of the 
sunspot number at each change. The phenomenological rule represented by Equation 
(\ref{eq:pop_law}) was found to hold to a very good approximation for the range $10<\Delta 
s<60$ over the last 14 cycles, although departures from the rule for small changes in sunspot 
number were noted. This departure may be attributed to the discrete nature of sunspot number, 
which means that the minimum sunspot number larger than zero is $11k$ (where $k$ is 
typically less than unity). This causes a discrete jump in the tabulated daily values of the 
international sunspot number from zero to seven (and implies that the average value of $k$ used 
by observers is $k=0.64$). Departures for large changes in sunspot number were also noted. 
This may be attributed to the finite size of the sunspot number over any cycle. Large negative 
changes in sunspot number are unlikely because the sunspot number is unlikely to have a 
sufficiently large value at any given time over a cycle to allow that change.

The distribution of a change $\Delta s$ on a given day is dependent on the value of sunspot 
number on that day. To model this we introduce transition probabilities
\begin{equation}
\label{eq:trans1}
p(s \rightarrow s'|s) = p(s',s)
\end{equation}
for changes from an initial sunspot number $s$ to a final sunspot number $s'$ on a day, given 
that the sunspot number is initially $s$. The function on the right hand side is a conditional 
distribution, and in Section \ref{subsection:transition} we consider a suitable functional form 
for this distribution. In Section \ref{subsection:relationship} we relate the chosen form of the 
conditional distribution $p(s',s)$ and \citeauthor{2012SoPh..276..351P}'s exponential 
distribution $f(\Delta s)$ of changes over complete cycles, and demonstrate how to estimate 
parameters from the data, for the proposed conditional distribution.

\subsection{The Form of the Conditional Distribution}
\label{subsection:transition}
\begin{figure}[ht!]
\centerline{\includegraphics[height=2.5in]{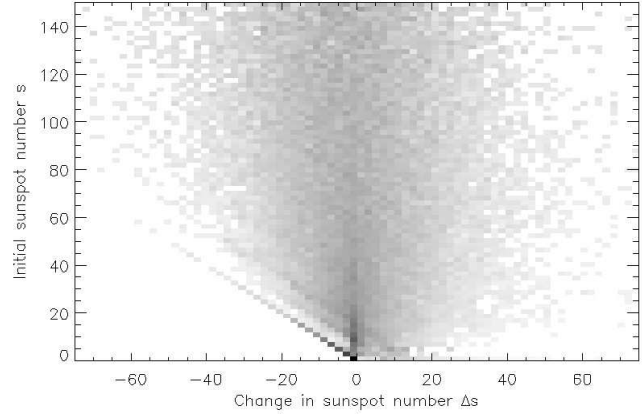}
}
\caption{%
Two dimensional histogram showing the fraction of days on which the sunspot number changes 
by $\Delta s$ (horizontal axis), given the initial value $s$ (vertical axis). The histogram has 
been normalised to represent an equal number of days at each initial sunspot number $s$.
}
\label{fig:2x}
\end{figure}

To gain insight into a suitable function form for the conditional distribution in Equation 
(\ref{eq:trans1}) we re-examine the data. Figure \ref{fig:2x} is a two-dimensional (2--D) 
histogram $N(\Delta s,s)$ of the number of days for which the sunspot number simultaneously 
has the value $s$ (enumerated along the vertical axis), and increases by $\Delta s$ to the next 
day (horizontal axis), for the NGDC data for 1815-2010. The bins are chosen to be of size two 
in $\Delta s$ and $s$, and the figure shows the normalised histogram
\begin{equation}
p_{i,j} = N(\Delta s_i,s_j)/\sum_i N(\Delta s_i,s_j),
\end{equation}
so that the greyscale density along each row shows the relative probability of a given change 
$\Delta s_i$, for the given initial sunspot number $s_j$. A nonlinear scaling is applied to the 
greyscale density, to better show the bins with small numbers of days. Figure \ref{fig:2x} 
illustrates the influence of the lower boundary $\Delta s = -s$ required by the non--negativity of 
sunspot number. The distribution is relatively symmetric about $\Delta s = 0$ for any given 
initial sunspot number $s$, except for an excess of days at the $\Delta s=-s$ boundary (which 
correspond to changes leading to a zero sunspot number), and an excess of days with no change 
({\it i.e.} $\Delta s = 0$).

Based on Figure \ref{fig:2x}, a suitable approximate model form for Equation (\ref{eq:trans1}) 
is a simple exponential distribution symmetric about $\Delta s=0$ for each value of $s$, with 
the same coefficient in the exponent for both negative and positive changes. The (asymmetric) 
non--negativity of sunspot number may be imposed by requiring that transitions producing a 
negative final sunspot number $(s'<0)$ lead to zero sunspot number $(s'=0)$ instead. This may 
be written as:
\begin{eqnarray}\label{eq:pop_law_cond}
p(s',s) & = & D\exp\left[-(s-s')/E\right]\mathbb{I}(s'<s) +
D\exp\left[(s-s')/E\right] \mathbb{I}(s'>s)\nonumber \\
& & + G\delta(s') + H\delta(s'-s),
\end{eqnarray}
where $D,\, E,\, F,\, G$, and $H$ are constants, and where the two terms involving delta 
functions describe the excess of days with changes leading to zero final sunspot number, and the 
excess of days on which the sunspot number does not change, respectively. The data show that 
for the observed changes in daily sunspot number for 1850--2011 the fraction of negative and 
positive jumps are approximately equal (42$\%$ and $41\%$ respectively), so we assume the 
same symmetry holds for each value of $s$ in Equation (\ref{eq:pop_law_cond}):
\begin{equation}
\int_{0}^{s}p(s',s)ds' = \int_{s}^{\infty}p(s',s)ds'.
\end{equation}
Normalising over all final sunspot numbers $s'$, {\it i.e.} requiring
\begin{equation}
\int_{0}^{\infty}p(s',s)ds' = 1
\end{equation}
leads to
\begin{equation}
 D = \frac{1-H}{2E}
\end{equation}
and
\begin{equation}
 G = \frac{1}{2}(1-H) {\rm e}^{-s/E}.
\end{equation}
The model conditional distribution (Equation (\ref{eq:pop_law_cond})) then has two 
parameters, $E$ and $H$. The parameter $E$ determines the size of changes on days when 
there is a change (a typical value, based on the data, is $E\approx 10$). The parameter $H$ is 
the probability of no change in daily sunspot number.

\subsection{Relating the Conditional Distribution and the \cite{2012SoPh..276..351P} 
Distribution, and Parameter Estimation}
\label{subsection:relationship}
The overall distribution of changes $f(\Delta s)$ may be calculated from the transitional 
distribution (Equation (\ref{eq:pop_law_cond})) by integrating over all starting values $s$, {\it 
i.e.} calculating
\begin{equation}
\label{eq:formula}
f(\Delta s) = L \int_{0}^{\infty}p(s+\Delta s,s)g(s)ds,
\end{equation}
where $L$ is a constant imposing normalisation over changes in sunspot number:
\begin{equation}
\label{eq:deltanorm}
\int_{-\infty}^{\infty}f(\Delta s) d \Delta s = 1,
\end{equation}
and $g(s)$ is the probability of an initial sunspot number $s$ in the observations. A suitable 
choice to approximately describe the distribution of sunspot number $g(s)$ over a complete 
solar cycle is an exponential \citep{2011ApJ...732....5N}\footnote{A noted in Section 
\ref{Section1}, the adherence of the overall sunspot number to an exponential distribution is 
only very approximate (by comparison with the observed distribution of changes in daily 
sunspot number, which follows the Laplace distribution quite strictly).}
\begin{equation}
g(s) = \frac{1}{\lambda}{\rm e}^{-s/\lambda},
\end{equation}
where the mean value of daily sunspot number (based on observations for 1850--2011) is 
$\lambda = 55$. Calculating the integral in Equation (\ref{eq:formula}) gives
\begin{eqnarray}
\label{eq:modelchanges}
L f(\Delta s) = &\frac{1-H}{2E}{\rm e}^{\Delta s/E}\left[1+\frac{E}{\lambda}e^{\Delta s
 /\lambda}\right] \mathbb{I}(\Delta s < 0) + \frac{1-H}{2E}{\rm e}^{-\Delta s/E}
 \mathbb{I}(\Delta s > 0) \nonumber \\
 & + H \delta (\Delta s )
\end{eqnarray}
and the normalisation by Equation (\ref{eq:deltanorm}) implies
\begin{equation}
L = 1 + \frac{E(1-H)}{2(\lambda+E)}.
\end{equation}
For the case of negative changes ($\Delta s<0$), Equation (\ref{eq:modelchanges}) contains a 
term $\lambda^{-1} E {\rm e}^{\Delta s/\lambda}$ which makes the distribution $f(\Delta 
s)$ asymmetric about $\Delta s=0$, and which is produced by changes leading to zero sunspot 
number. The characteristic size of this term is $E/\lambda\approx0.18$, suggesting that the 
asymmetry in the distribution in $\Delta s$ produced by the lower boundary $\Delta s = -s$ is 
not a strong effect. In the absence of the extra term the normalisation constant is $L=1$, and 
based on the data we find $L\approx 1.05$. Neglecting the extra term and setting $L=1$, the 
distribution of changes implied by the conditional distribution (Equation 
(\ref{eq:pop_law_cond})) is
\begin{eqnarray}
\label{eq:approxmodelchanges}
f(\Delta s) = &\frac{1-H}{2E}{\rm e}^{\Delta s/E}\mathbb{I}(\Delta s < 0) +
\frac{1-H}{2E}{\rm e}^{-\Delta s/E} \mathbb{I}(\Delta s > 0) + H \delta (\Delta s )
\end{eqnarray}
which is the same functional form as Equation (\ref{eq:pop_law}), the 
\cite{2012SoPh..276..351P} distribution of changes.

The correspondence between the conditional distribution (Equation (\ref{eq:pop_law_cond})) 
and the overall distribution of changes (Equation (\ref{eq:approxmodelchanges})) allows the 
parameters $E$ and $H$ of the conditional distribution to be estimated from the daily changes 
$\Delta s$ over a cycle using maximum likelihood. Specifically, the exponential coefficient 
$B$ (see Equation (\ref{eq:pop_law_cond})) estimated for the overall distribution may be 
taken as the estimate for coefficient $E$ in the conditional distribution, and the fraction of days 
$C$ with no change in sunspot number (see Equation (\ref{eq:pop_law_cond})) in the overall 
distribution may be taken as the estimate of the corresponding parameter $H$ in the conditional 
distribution.

\section{Simulating Sunspot Numbers}
\label{Section:Section3}
In this section we apply the conditional distribution (Equation (\ref{eq:pop_law_cond})) in a 
Monte Carlo simulation of daily sunspot numbers $s_i=s(t_i)$, where $t_i$ refers to a day. The 
daily random variation in sunspot number is described by the stochastic differential equation 
(stochastic DE)
\begin{equation}\label{eq:sde_trans}
\frac{ds}{dt}=\sum_{i=1}^{N}\Delta s_i,
\end{equation}
where $N$ is the number of days and the daily change $\Delta s_i$ is generated by sampling 
from the distribution
\begin{equation}
p(s'_i,s_i) = p(s_i + \Delta s_i,s_i)
\end{equation}
with $p(s',s)$ given by Equation~(\ref{eq:pop_law_cond}).

To solve the stochastic DE (Equation~(\ref{eq:sde_trans})) we need to sample from the 
conditional distribution $p(s',s)$ given by Equation (\ref{eq:pop_law_cond}), with our 
maximum likelihood estimate of the parameters $E$ and $H$. This is achieved as follows. For a 
given initial sunspot number $s$ and the estimates of $E$ and $H$ we generate a final sunspot 
number $s'=s+\Delta s$ by generating an exponential random variable $\Delta s$ with 
parameter $E$ {\it i.e.} a random deviate $\Delta s$ distributed according to $\sim {\rm 
e}^{-\Delta s/E}$. A random variable $u$ which is uniformly distribution on $(0,1)$ is also 
generated, and then the final change $\Delta s$ is calculated according to the rule:
\begin{itemize}
 \item if $u < 0.5(1-H)$, then $\Delta s = -\Delta s$;
 \item if $u > 0.5(1+H)$, then $\Delta s = \Delta s$;
 \item otherwise $\Delta s = 0$.
\end{itemize}
This procedure assigns no change in sunspot number ({\it i.e.} $\Delta s=0$) with probability 
$H$, and the remaining changes are exponentially distributed over positive and negative 
$\Delta s$ with equal total probability. The mean absolute size of changes (on days when 
$\Delta s \neq 0$) is $E$. Finally, to prevent the final sunspot number $s' = s+\Delta s$ from 
being negative, if $s+\Delta s<0$, then we take $\Delta s = -s$.

This procedure assigns values correctly according to Equation (\ref{eq:pop_law_cond}). 
Equation~(\ref{eq:sde_trans}) is solved for a given initial sunspot number $s_0$ at time 
$t_0$ by generating a sequence of transitions $\Delta s_i$ (with $i=1,2,...,N$) according to this 
recipe, and adding these successively to $s_0$.

This model accounts for the stochastic variation in sunspot number according to the conditional 
distribution given by Equation~(\ref{eq:pop_law_cond}), but it does not account for the secular 
or long time-scale variation of the sunspot number over a solar cycle (the ``shape'' of the cycle). 
Recently \citet{2011ApJ...732....5N} presented a method for modelling the solar cycle variation 
in a general Fokker-Planck description of stochastic variation in sunspot number, and the same 
procedure is applied here. A term may be added to the stochastic DE (Equation 
(\ref{eq:sde_trans})) causing the fluctuating sunspot number to return to a prescribed time 
evolution $\theta (t)$:
\begin{equation}\label{eq:sde_sec_trans}
\frac{ds}{dt}= \kappa \left[ \theta(t) - s \right]
+\sum_{i=1}^{N}\Delta s_i.
\end{equation}
The function $\theta(t)$ is referred to as the driver function, and factor $\kappa$ is the rate at 
which sunspot number $s$ returns to the value specified by the driver function. The two terms 
on the right hand side of Equation~(\ref{eq:sde_sec_trans}) are deterministic and stochastic 
terms, respectively. Equation~(\ref{eq:sde_sec_trans}) is solved for a given initial sunspot 
number $s_0$ by adding daily stochastic transitions $\Delta s_i$ in the same way as for 
Equation~(\ref{eq:sde_trans}). In between the transitions the sunspot number is evolved 
according to Equation~(\ref{eq:sde_sec_trans}) with just the deterministic term included. The 
solution to the differential equation with just the deterministic term is
\begin{equation}\label{eq:sde_sec_sol}
s^*(t)={\rm e}^{-\kappa (t-t_i)} \left\{ s_i + \kappa \int_{t_i}^{t}
\theta(t') {\rm e}^{\kappa t'}dt' \right\},
\end{equation}
where $s_i=s(t_i)$ is the value of the sunspot number on the most recent day, and 
$t_i<t<t_{i+1}$.

To summarise, the procedure for simulating the sunspot number evolution for day $i+1$, given 
the sunspot number $s_i$ on day $i$, is to evaluate a deterministic value for the sunspot number 
$s^*(t_{i+1})$ using Equation~(\ref{eq:sde_sec_sol}), to generate a random change $\Delta 
s_{i+1}$ using Equation~(\ref{eq:pop_law_cond}) with initial sunspot number 
$s=s^*(t_{i+1})$, and then the sunspot number on day $i+1$ is $s_{i+1}=s^*(t_{i+1})+\Delta 
s_{i+1}$.

The driver function $\theta (t)$ in the deterministic term in Equation (\ref{eq:sde_sec_trans}) 
represents the functional form of the solar cycle variation in sunspot number, {\it i.e.} the shape 
of a cycle. The driver function describes basic empirical features of a solar cycle, such as the 
time taken to reach maximum, the size of the maximum, and so on. Additionally, 
$\theta(t)$ may account for detailed features of the shape of a cycles such as the Gnevyshev 
Gap \citep{1967SoPh....1..107G}. Here we use a function introduced by 
\citeauthor{1994SoPh..151..177H} (\citeyear{1994SoPh..151..177H}, hereafter HWR94):
\begin{equation}
\label{eq:hath}
\theta(t) = \frac{a \left( t-t_0 \right)^3}{\exp\left[ -(t-t_0)^2 /b^2
\right] - c},
\end{equation}
where $t_0$ is the start time for a cycle, and $a$, $b$, and $c$ represent the cycle amplitude, 
period, and asymmetry, respectively. Equation (\ref{eq:hath}) was used by 
\cite{2012SoPh..276..363N} to investigate daily variation in sunspot number, and to forecast 
sunspot number and solar cycles. A statistical procedure for calculating estimates of the 
parameters $a$, $b$, $c$, and $\kappa$ from daily sunspot data, and the values of the estimates 
for cycles 11--23, was given in \cite{2012SoPh..276..363N}. Here we re--use these parameter 
estimates describing the shapes of the cycle to simulate three recent solar cycles (21, 22, and 
23).

\begin{figure}[ht!]
\centerline{\includegraphics[height=3.0in,width=4.0in]{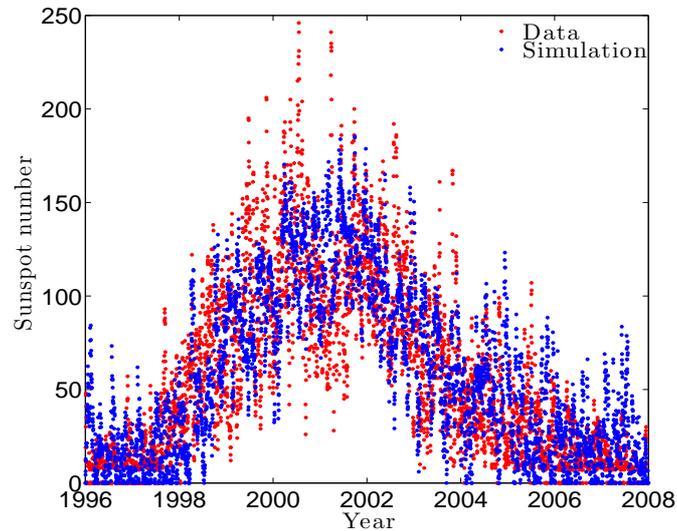}
}
\caption{%
The observed daily sunspot numbers over cycle 23 (red points), and a simulation of the sunspot 
numbers using the procedure outlined in Section \ref{Section:Section3}. The parameters for the 
modelling of the shape of the cycle are taken from \cite{2012SoPh..276..363N}.
}
\label{fig:f4}
\end{figure}

Figure \ref{fig:f4} shows the daily sunspot numbers over cycle 23, for the years 1996 to 2008 
(red points), and our simulation of sunspot numbers for this cycle based on Equation 
(\ref{eq:sde_sec_trans}) (blue points). The HWR94 driver function, Equation (\ref{eq:hath}), 
enforces the secular variation in the solar cycle, with the parameter values $a=7.82\times 
10^{-8}$, $b=1514$, $c=0.222$, and $\kappa=0.086$ (taken from Table I in 
\cite{2012SoPh..276..363N}). We also use the estimates $E=8.78$ and $H=0.149$ for the 
conditional distribution (Equation (\ref{eq:pop_law_cond})), which are maximum likelihood 
values on the changes in daily sunspot data for cycle 23, as discussed in Section 
\ref{Section:Section2}. For the HWR94 driver function it is not possible to solve Equation 
(\ref{eq:sde_sec_sol}) analytically, and instead we integrate
\begin{equation}
\frac{ds}{dt} = \kappa \left[ \theta(t) - s \right]
\end{equation}
numerically using a fourth order Runge-Kutta scheme \citep{1992nrca.book.....P}.

\begin{figure}[ht!]
\centerline{\includegraphics[height=4.5in,width=4.0in]{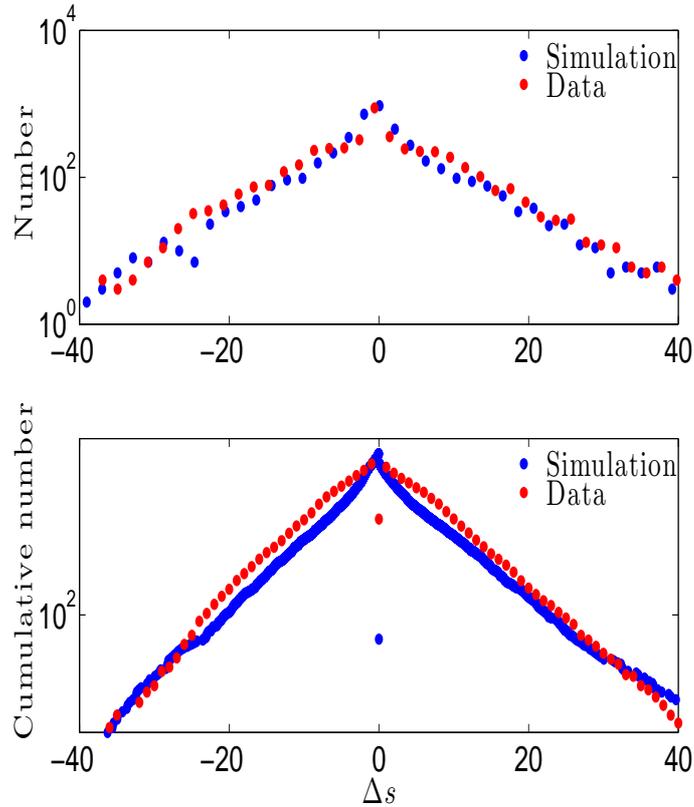}
}
\caption{%
Histograms of the observed daily changes in the sunspot number for cycle 23 (see Figure 
\ref{fig:f4}), and for our simulation of the sunspot numbers. The red points show the data and 
the blue points the simulation. The upper panel shows the numbers of days with the given 
change in sunspot number and the lower panel shows the corresponding cumulative distribution.
}
\label{fig:f5}
\end{figure}

Figure \ref{fig:f5} shows the daily changes in sunspot number for cycle 23 (red points), and the 
corresponding changes in our simulation (blue points). The upper panel of Figure \ref{fig:f5}
shows the distribution of changes and the lower panel shows the cumulative distribution in the 
same format as Figure \ref{fig:f1}. Figure \ref{fig:f5} confirms that the simulation of daily 
sunspot number over cycle 23 based on the stochastic DE (Equation~(\ref{eq:sde_sec_trans})) 
and the conditional distribution (Equation (\ref{eq:pop_law_cond})), together with the HWR94 
model for the shape of the solar cycle, generates a distribution of changes $f(\Delta s)$ over the 
cycle that closely resembles the exponential form identified by \cite{2012SoPh..276..351P}.

\begin{figure}[ht!]
\centerline{\includegraphics[height=4.5in,width=4.0in]{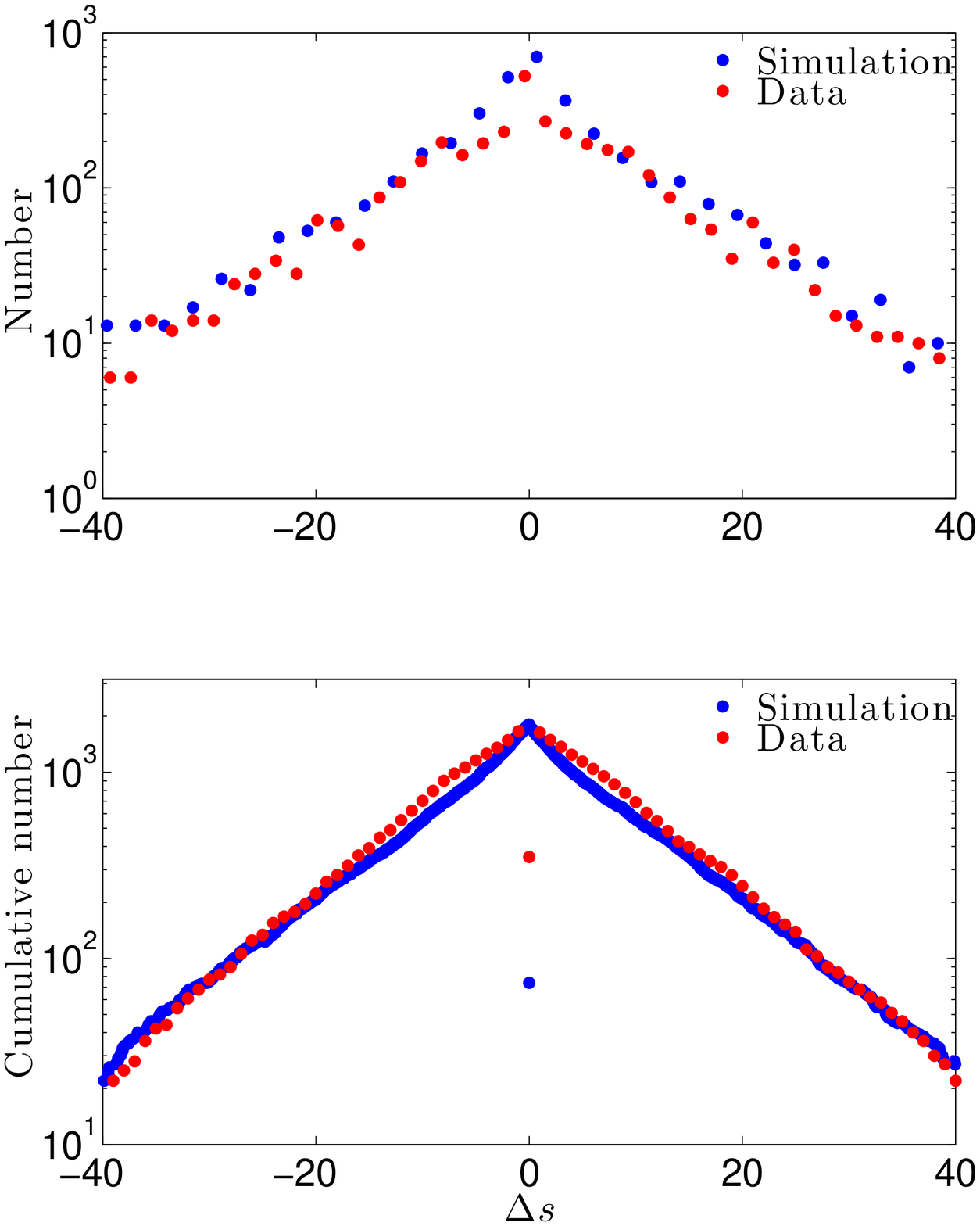}
}
\caption{%
Changes in daily sunspot number for solar cycle 22 (red points) and for our simulation of the 
cycle (blue points). The upper panel shows the numbers of days with the given change, and the 
lower panel is the corresponding cumulative distribution.
}
\label{fig:f6}
\end{figure}

Figure \ref{fig:f6} presents the results of the simulation procedure for cycle 22 (years 1986 to 
1996) in the same format as Figure \ref{fig:f5}. The parameter estimates for the HWR94 driver 
function are $a=14.1\times 10^{-8}$, $b=1368$, $c=0.33$, and $\kappa=0.073$, again taken 
from Table I in \cite{2012SoPh..276..363N}. The estimates $E=10.4$ and $H=0.096$ are used 
for the parameters in the conditional distribution Equation (\ref{eq:pop_law_cond}), based on 
maximum likelihood applied to the daily data for cycle 22.

\begin{figure}[ht!]
\centerline{\includegraphics[height=4.5in,width=4.0in]{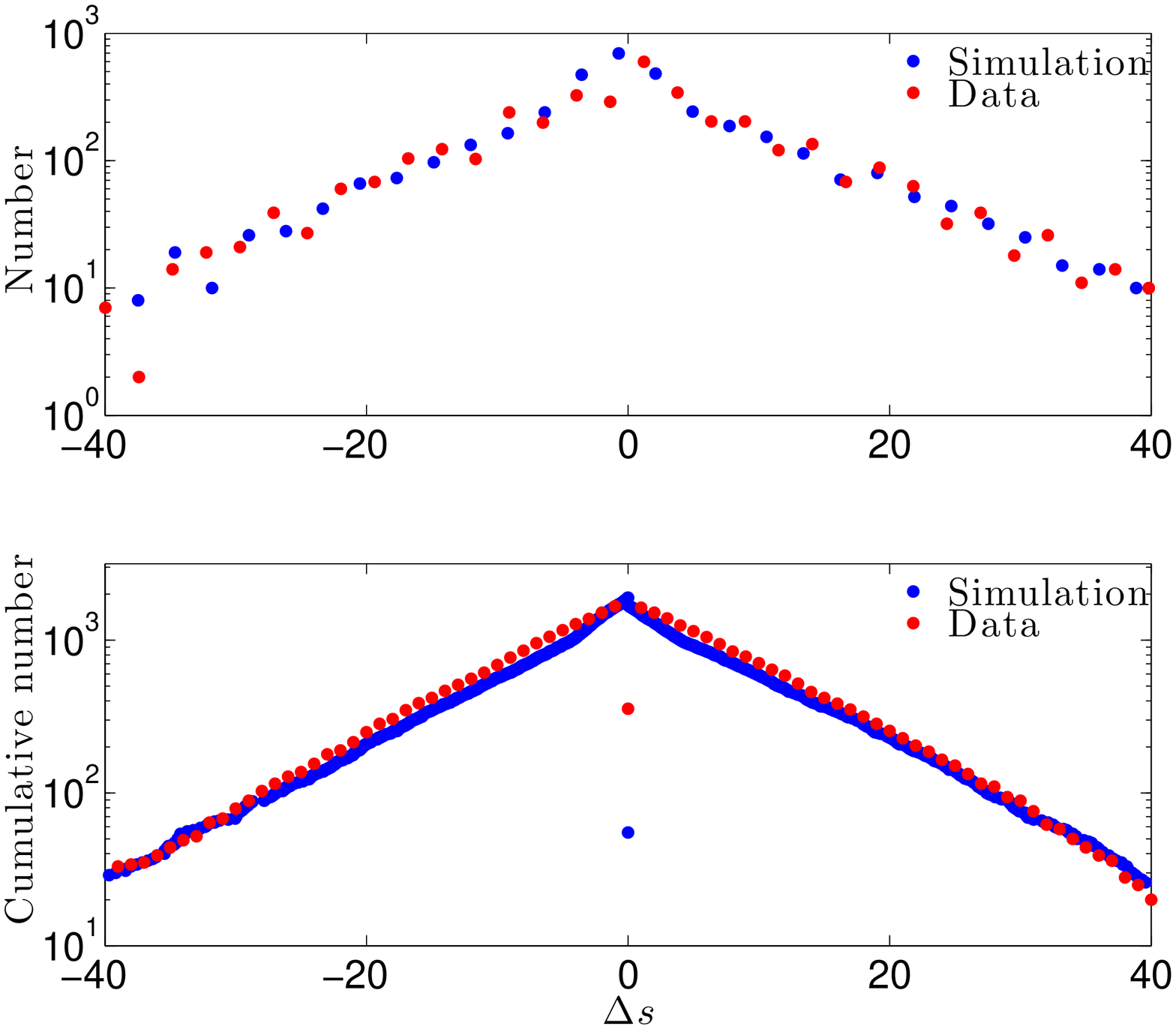}
}
\caption{%
Changes in daily sunspot numbers for solar cycle 21 (red points) and for our simulation of the 
cycle (blue points). The upper panel shows the number of days with the given change, and the 
lower panel is the corresponding cumulative distribution.
}
\label{fig:f7}
\end{figure}

Figure \ref{fig:f7} presents the results of the simulation procedure for cycle 21 (years 1976 to 
1986), again in the same format as Figure \ref{fig:f5}. The parameter estimates for the HWR94 
driver function are $a=12.2\times 10^{-8}$, $b=1414$, $c=0.490$, and $\kappa=0.073$, again 
taken from Table I in \cite{2012SoPh..276..363N}. The maximum likelihood estimates 
$E=10.6$ and $H=0.098$ are used for the parameters in the conditional distribution (Equation 
(\ref{eq:pop_law_cond})).

Figures \ref{fig:f6} and \ref{fig:f7} confirm that the simulation procedure succeeds in 
reproducing the phenomenological exponential rule for daily changes in sunspot number when 
applied to cycles 22 and 21, respectively.

\section{Discussion}
\label{Section:Section5}
This paper establishes that the observed Laplace, or double exponential distribution of changes 
$\Delta s$ in daily sunspot number $s$ (for days on which the sunspot number does change) 
recently identified by \cite{2012SoPh..276..351P}, is due to the evolution of observed sunspot 
groups ({\it i.e.} group formation, spot splitting, spot/group decay) rather than being due to the 
artificial variation caused by groups rotating onto and off the visible disc. The implication is 
that the distribution has a physical basis. Sunspot emergence, evolution, and eventual decay 
produces daily changes in sunspot number which may be positive or negative, and changes of 
this kind in separate active regions may add or cancel. The sum of these daily changes, 
remarkably, produces a simple Laplace distribution, with a marked excess of days with no 
change in sunspot number.

In this paper we show also how to simulate daily sunspot number via a Monte Carlo method, 
using a conditional distribution based on the exponential rule together with a model for the solar 
cycle variation in sunspot number. The conditional distribution $p(s',s)$ introduced describes 
the probability of a change from a current sunspot number $s$ to a value $s'=s+\Delta s$ in one 
day, given the initial sunspot number, and ensures that $s' \geq 0$. The simulation procedure 
involves calculating a secular or deterministic change in sunspot number due to the underlying 
solar cycle, and then adding a random change in sunspot number according to the conditional 
distribution. The Monte Carlo method is demonstrated in application to three recent solar cycles 
(cycles 21, 22, and 23). The simulated sunspot numbers exhibit a distribution of changes 
$f(\Delta s)$ over each cycle that closely reproduces the Laplace distribution identified by 
\cite{2012SoPh..276..351P}.

It is interesting to consider possible explanations for the observed Laplace distribution. The 
origin of the surface changes in sunspot number described by the rule are changes in the 
structure of the subphotospheric magnetic fields, which are not directly amenable to observation 
\citep{ThomasEtAl1991}, although local helioseismology is beginning to provide some 
insights \citep{2005LRSP....2....6G}. In the absence of detailed physical models for the surface 
changes provided by this field evolution, it may be possible to construct statistical models for 
daily changes in sunspot number based on simple statistical descriptions of spot and group 
evolution, for given numbers of spots and groups. For example, probabilities could be assigned 
to given spots or groups increasing or decreasing their number in a day. Such a description may 
be modelled by a type of birth--death process, which have been used in the natural and social 
sciences ({\it e.g.} see for example \opencite{Gillespie1992}). It may also be possible to use 
other known statistical rules for the distribution and evolution of spot groups, for example the 
log--normal distribution of spot areas (\citeauthor{1988ApJ...327..451B} 
\citeyear{1988ApJ...327..451B}), and various rules for the decay rate of sunspot area (in area 
per unit time) per sunspot within a group (see \opencite{2003A&ARv..11..153S}). Given the 
complexity of the combinations implied by this consideration of possible modelling, we note 
again how remarkable it is that a simple double exponential form arises. We leave the 
development of a detailed model to a future investigation.

\begin{acks}
P.~N. gratefully acknowledges a University of Sydney Postgraduate Scholarship.
\end{acks}

% FIGURES

\end{article}
\end{document}